\documentclass[twocolumn,showpacs,preprintnumbers,amsmath,amssymb,superscriptaddress]{revtex4}

\usepackage{graphicx}
\usepackage{dcolumn}
\usepackage{bm}
\usepackage{amsmath}
\usepackage{color}

\begin{document}

\preprint{APS/123-QED}

\title{Coherent Magnetism and Spin-Orbit Interaction in Garnet Films \\ Probed with Femtosecond Magneto-Optical Four Wave Mixing  \\}

\author{M. Barthelemy}%
    \affiliation{Institut de Physique et Chimie des Mat\'eriaux de Strasbourg, (UMR 7504 CNRS-UdS), Strasbourg 67034, France}%
\author{M. Vomir}%
    \affiliation{Institut de Physique et Chimie des Mat\'eriaux de Strasbourg, (UMR 7504 CNRS-UdS), Strasbourg 67034, France}%
\author{M. Sanches Piaia}%
    \affiliation{Institut de Physique et Chimie des Mat\'eriaux de Strasbourg, (UMR 7504 CNRS-UdS), Strasbourg 67034, France}%
\author{H. Vonesch}%
    \affiliation{Institut de Physique et Chimie des Mat\'eriaux de Strasbourg, (UMR 7504 CNRS-UdS), Strasbourg 67034, France}%
\author{P. Molho}%
    \affiliation{Institut N\'eel, CNRS - Universit\'e Joseph Fourrier, 25 avenue des Martyrs, b\^atiment K, BP 166 38042 Grenoble cedex 9, France}%
\author{B. Barbara}%
    \affiliation{Institut N\'eel, CNRS - Universit\'e Joseph Fourrier, 25 avenue des Martyrs, b\^atiment K, BP 166 38042 Grenoble cedex 9, France}%
\author{J.-Y. Bigot}%
    \affiliation{Institut de Physique et Chimie des Mat\'eriaux de Strasbourg, (UMR 7504 CNRS-UdS), Strasbourg 67034, France}%
    \email{bigot@ipcms.u-strasbg.fr}

\date{\today}

\begin{abstract}
We report about the dephasing of the spins states in a Garnet film excited by femtosecond laser pulses. It is shown that magneto-optical four-wave-mixing signals are efficiently generated and controlled with an external static magnetic field. The self-diffraction of either two or three femtosecond pulses allows us determining independently the coherent response time ($T_2 < \text{50 fs}$) and the population relaxation, which occurs in several steps. After the laser induced demagnetization, the spin-lattice relaxation is the dominant mechanism ($T_{1,spin-lat} < \text{1.35 ps}$) and the precession of the magnetization takes place at later times ($T_{1,prec} < \text{150 ps}$). The coherent response of the spins states is modeled with a multi-level hydrogen-like system, taking into account the fundamental relativistic quantum dynamics of the spins. Our experimental approach is general and opens the way to investigating quantum magnetism in ordered magnetic systems useful for applications in information and communications technologies.
\end{abstract}


\maketitle
To improve the speed of writing and reading information in recording media one can use femtosecond laser pulses instead of pulsed magnetic fields. Ideally one would like to manipulate the magnetization of a material which has a memory (ferromagnetic), by modifying the polarization of the light. For example, the magnetization of GdFeCo ferrimagnetic thin films can be reversed locally with circularly polarized pulses \cite{Stanciu2007}, a process occurring within a few tens of picoseconds \cite{Vahaplar2009}. From a quantum point of view, both the coherence and spins populations are modified when the spins interact with the optical field via the spin-orbit Hamiltonian. Therefore, when performing optical transitions in a ferro or ferrimagnetic material one has to consider both processes. We could show earlier that the coherent photon-spin interaction occurs in ferromagnetic metallic films by studying their time resolved magneto-optical response in a pump-probe geometry \cite{JYB2009}.

While it is well known that the spins populations dynamics leads to a demagnetization \cite{Beaurepaire1996,Hohlfeld1997,Aeschlimann1997}, or may even induce a magnetic ordering \cite{Perakis2013}, little is known about the coherence of the magnetic states. Population effects are easily understandable as they can be modeled with raising electron and spin temperatures leading to an ultrafast demagnetization. Typically this process occurs within  100 fs during the thermalization of the spins at the Fermi level of the metal \cite{Guidoni2002,Schmidt2005,Radu2009} and recently it has been shown \cite{Boeglin2010} to be driven by the interaction between the orbital and spin angular momentum of the electrons. Additionally, the spin-phonon interaction either with a thermal \cite{Koopmans2010} or non-thermal \cite{Oppeneer2013} electron distribution can be partially involved. In contrast, coherent processes are directly related to the polarization of the electromagnetic field. For example, the precession of the magnetization in antiferromagnetic \cite{Kimel2004} or ferrimagnetic \cite{Hansteen2005} thin films can be controlled by using either  $\sigma^+$ or $\sigma^-$ circularly polarized light pulses in a pump probe configuration. Regarding the coherent coupling between the laser field and the spins, it involves the spin-orbit coupling \cite{Zhang2000,Zhang2009, Vonesch2012} and, in its most general description, it includes the potential of the nuclei as well as the one of the electromagnetic field itself \cite{JYB2010, Hinschberger2012}.

In the present work, we have studied the coherent magneto-optical response independently of the population dynamics. Towards that goal, we measured magneto-optical four wave mixing signals (MO-FWM) observed in a Bismuth doped Garnet film with perpendicular magneto-crystalline anisotropy. We are able to distinguish between the dephasing dynamics ($T_2 < \text{50 fs}$) of the coherent response and the lifetime of the spins population. Two different physical mechanisms are involved. On one hand the MO-FWM emission finds its origin in the large coherent coupling between the laser field and the spins, mediated by the spin-orbit interaction. It can be accounted for by using a spin-photon interaction Hamiltonian, including the relativistic corrections at first order in $1/c^2$. On the second hand, the dynamics of the spins population is a multi-step process dominated by the initial electron-lattice relaxation ($T_{1,spin-lat} < \text{1.35 ps}$) as well as the precession and damping of the magnetization ($T_{1,prec} < \text{150 ps}$).

\begin{figure}
\includegraphics[scale=0.3]{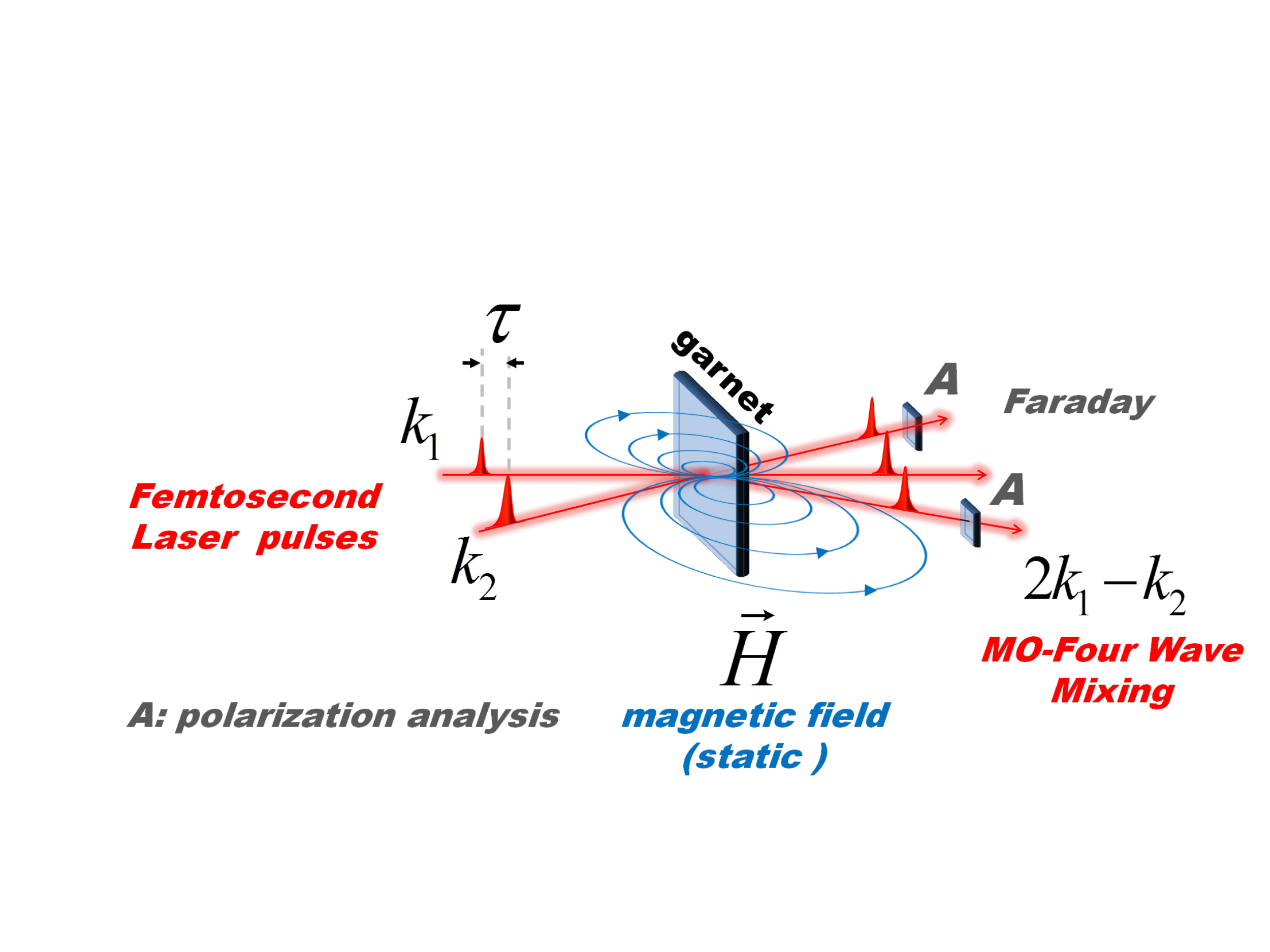}
\caption{\label{fig1}Sketch of the Femtosecond Magneto-optical Four Wave Mixing configuration for two pulses.}
\end{figure}

\begin{figure}
\includegraphics[scale=0.3]{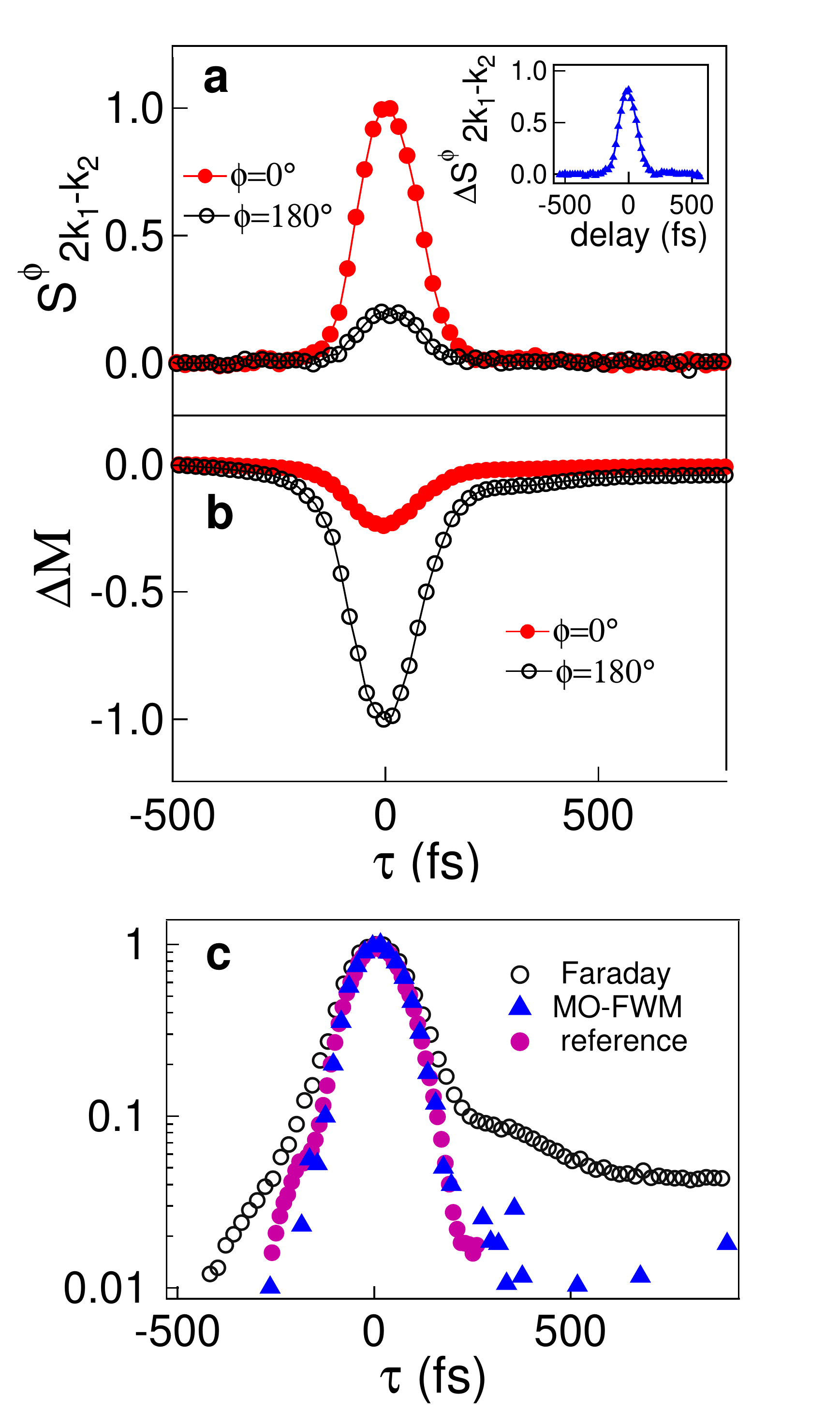}
\caption{\label{fig2}Magneto-optical Four-Wave Mixing and Faraday signals. a) MO-FWM signals $S_{2k_1-k_2}^\phi(\tau)$ for the two orientations  $\phi = 0^\circ (180^\circ)$ of the magnetic field, normalized to the signal obtained for $\phi = 0^\circ$. The inset shows the difference between the two signals $\Delta S_{2k_1-k_2}^\phi(\tau) = S_{2k_1-k_2}^\phi(\tau) - S_{2k_1-k_2}^{\pi - \phi}(\tau)$. b) Faraday signal measured in the direction $\mathbf{k}_2$. c) Comparison with a reference signal obtained in a thin film of $TiO_2$.}
\end{figure}

In a first set of experiments, we measured MO-FWM signals generated by two linearly polarized femtosecond pulses as sketched in figure 1. In this self-diffraction configuration, the two frequency degenerate pulses propagate in the directions $\mathbf{k}_1$ and $\mathbf{k}_2$. The magneto-optical signal $S_{2k_1-k_2}^\phi(\tau)$ is observed in the direction $2 \mathbf{k}_1 - \mathbf{k}_2$ as a function of the delay $\tau$ between the two pulses to obtain the information on the dephasing of the spin states. The orientation $\phi$ of the external static magnetic field $\mathbf{H}$ is also varied to obtain the MO-FWM signal for opposite spins directions. $\phi = 0^\circ (180^\circ)$ corresponds to the field +H (-H) perpendicular to the sample plane. The polarization of the MO-FWM signal is analyzed at an angle $\alpha$ close to the crossed polarization of the incident beam as performed in the analysis of magneto-optical Kerr or Faraday signals\cite{JYB2004}. Simultaneously, the magneto-optical Faraday signal $\Delta M(\tau)$ is measured in the direction $\mathbf{k}_2$. The pulses are issued from an amplified Titanium Sapphire laser functioning at 5 kHz. They both have durations of 120 fs and a central wavelength of 794 nm. They are focused with a spot diameter of 100 $\mu$m onto the sample where they spatially overlap. The delay $\tau$ between the two pulses is varied with a stepper motor delay line. The densities of energy of the pump and probe pulses are: $I_{k_1} = \text{1.45 mJcm}^{-2}$ and $I_{k_2} = \text{1.27 mJcm}^{-2}$. The sample is a 7 $\mu$m thick $\text{(GdTmPrBi)}_3\text{(FeGa)}_5\text{O}_{12}$  Garnet deposited on a GGG $\text{(Gd}_3\text{Ga}_5\text{O}_{12})$ substrate (0.5 mm thick) using the liquid phase epitaxy\cite{Ferrand1999}. It has a perpendicular magneto-crystalline anisotropy and a weak ferrimagnetic remanence at room temperature.

In a second type of experiments, we measured MO-FWM signals generated by three femtosecond pulses as sketched in the inset of figure 3a. In that case, the magneto-optical signal $S_{k_1-k_2+k_3}^\phi(\tau,T)$ is observed in the direction $\mathbf{k}_1 - \mathbf{k}_2 + \mathbf{k}_3$ either as a function of the delay $\tau$ between pulses $\mathbf{k}_1$ and $\mathbf{k}_2$ or as a function of the delay $T$ between pulses $\mathbf{k}_1$ and $\mathbf{k}_3$. Like for FWM signals generated by electronic populations in molecular systems\cite{Weiner1985}, for a fixed $T=T_0$, $S_{k_1-k_2+k_3}^\phi(\tau,T_0)$ corresponds to the dephasing dynamics with dephasing time $T_2$, while for a fixed delay $\tau = \tau_0$, $S_{k_1-k_2+k_3}^\phi(\tau_0,T)$ corresponds to the population dynamics with relaxation lifetime $T_1$. Note that for the three beams measurements, performed on a different laser system, the pulse duration is shorter (50 fs instead of 120 fs) and in that case the analysis of the polarization rotation is made with a bridge composed of a half-wave plate, a Wollaston prism and two photo-diodes.

Figure 2a) displays the magneto-optical four wave mixing signals $S_{2k_1-k_2}^\phi(\tau)$ for opposite directions  $\phi = 0^\circ$ and $180^\circ$ of $\mathbf{H}$. The difference between the two signals $\Delta S_{2k_1-k_2}^\phi(\tau)$ is shown in the inset. The time dependent magneto-optical Faraday signal $\Delta M(\tau)$ is shown in figure 2b). Both signals are compared on a logarithmic scale in figure 2c) (Faraday: opened circles, MO-FWM: closed circles) together with a reference FWM signal $S_{ref}(\tau)$ generated in a $TiO_2$ transparent crystal which is non-magnetic (FWM in $TiO_2$: crosses). The temporal variation of the Faraday signal contains two components. One corresponds to the coherent photon-spin interaction, while the second one is due to the recovery of the magnetization modulus initially reduced by the laser pulse $\mathbf{k}_1$. The mechanism involved is the spin-lattice interaction and has already been studied in many different materials using time resolved magneto-optical Faraday or Kerr configurations\cite{Beaurepaire1996,Hohlfeld1997,Guidoni2002}. It can be described with a three-temperature model representing the dynamics of the charges, the spins and the lattice\cite{JYB2001}.

In contrast, the magneto-optical FWM signal is purely coherent and its overall behavior is the same as the reference signal. They are both shorter than the pulse duration. The MO-FWM signal is quadratic as it is proportional to $I_{k_1}^2 I_{k_2}$, while the pump-probe magneto-optical Faraday signal is proportional to $I_{k_1} I_{k_2}$. The efficiency $\eta_{MO-FWM}$ of the MO-FWM emission is very large. By analogy with the third order nonlinear magneto-optical susceptibility, we define it as: $\eta_{MO-FWM} = \frac{1}{I_{k_1}}\sqrt{\frac{I_{FWM}}{I_{k_2}}}\frac{4n^2(\omega)\epsilon_0c^2}{d\omega}H_S$, ($\epsilon_0$: permittivity of vacuum, $n$: linear refractive index at frequency $\omega(\lambda = \text{794 nm})$, $d$: film thickness and $H_S$: applied magnetic field at saturation. With pulses duration of 120 fs focused on a spot diameter of (100 $\pm$ 25) $\mu$m we obtain: $\eta_{MO-FWM} = (2.1 \pm 0.5)\times10^{-18} \text{AmV}^{-2}$ using:  $I_{k_1} = 1.21\times10^{14}\text{Wm}^{-2},  I_{k_2} = 1.06\times10^{14} \text{Wm}^{-2} ; n_{\text{800 nm}} = 1,7 ; d = 7\times10^{-6} \text{m}, I_{FWM} = 4\times10^7 \text{Wm}^{-2} \text{ and } H_S=8\times10^2 \text{Am}^{-1}$.

The large MO-FWM emission $S_{2k_1-k_2}^\phi(\tau)$ is due to the coherent coupling between the photons and the spins. Similarly to the coherent magneto-optical response that has been reported in ferromagnetic metallic thin films \cite{JYB2009} the spin-orbit interaction plays a major role in the time dependent nonlinear magneto-optical response. We have developed a model consisting of a multi-level system interacting with a laser field \cite{Vonesch2012}. It takes into account the photon-spin interaction, including spin-flip processes due to the coupling of the laser field with the orbital and spin angular momenta. The Hamiltonian of interaction is: $H_{int} = \mathbf{A}_L(\frac{e}{m}\mathbf{p} + \frac{e^2}{m}\mathbf{A}_M - \frac{e^2}{2m^2}\mathbf{S} \wedge \mathbf{E_{ion}})$ with $\mathbf{A}_M$ and $\mathbf{A}_L$ being the magnetic and laser potential vectors, $\mathbf{S}$ the spin vector and $\mathbf{E}_{ion}$ the field of the ions, $\mathbf{p}$ the momentum of the electrons.

Using the density matrix \cite{Mukamel1995}, we have solved the Liouville equation for our hygrogen-like system, taking into account the time ordering of the pulses. We find the contributions to the dephasing and populations of spins states in the magneto-optical signals. For the MO-FWM signal, the temporal sequence of pulses which gives rise to a diffracted signal in the direction $2\mathbf{k}_1 - \mathbf{k}_2$ is: $E_2^*(t-\tau)E_1(t) E_1(t)$. The corresponding dephasing time $T_2$ is shorter than the pulse duration and cannot be resolved here. For the Magneto-Optical Faraday signal $\Delta M(\tau)$, observed in the direction $\mathbf{k}_2$, three contributions are present. Two of them are coherent: the "polarization free decay" which corresponds to the temporal sequence of the fields: $E_1(t)E_2^*(t-\tau)E_1^*(t)$ and the "pump perturbed free decay" corresponding to the temporal sequence: $E_2^*(t-\tau)E_1(t) E_1^*(t)$. They both relax with a dephasing time $T_2$. Just like the MO-FWM signal, $T_2$ being shorter than the laser pulse, we simply have a convolution of the three fields. Nevertheless these two contributions are the origin of the very large coherent MO-Faraday signal near zero delay time $\tau$ displayed in figure 2 (2b) and 2c) opened circles). The third temporal sequence is: $E_1(t)E_1^*(t)E_2^*(t-\tau)$. This term corresponds to the magnetization dynamics which relaxes with several lifetimes $T_1$. It basically contains all the "thermal" processes usually identified in the spins population dynamics. This population contribution is absent in the MO-FWM which is detected in a different direction. Note that in the case of metallic systems, like Ni and $CoPt_3$ thin films, the coherent contribution is also present \cite{JYB2009}. It is however weaker than the spins populations as most of the femtosecond Faraday signal is due to the thermalization of the spins in the case of transition metals.

\begin{figure}
\includegraphics[scale=0.3]{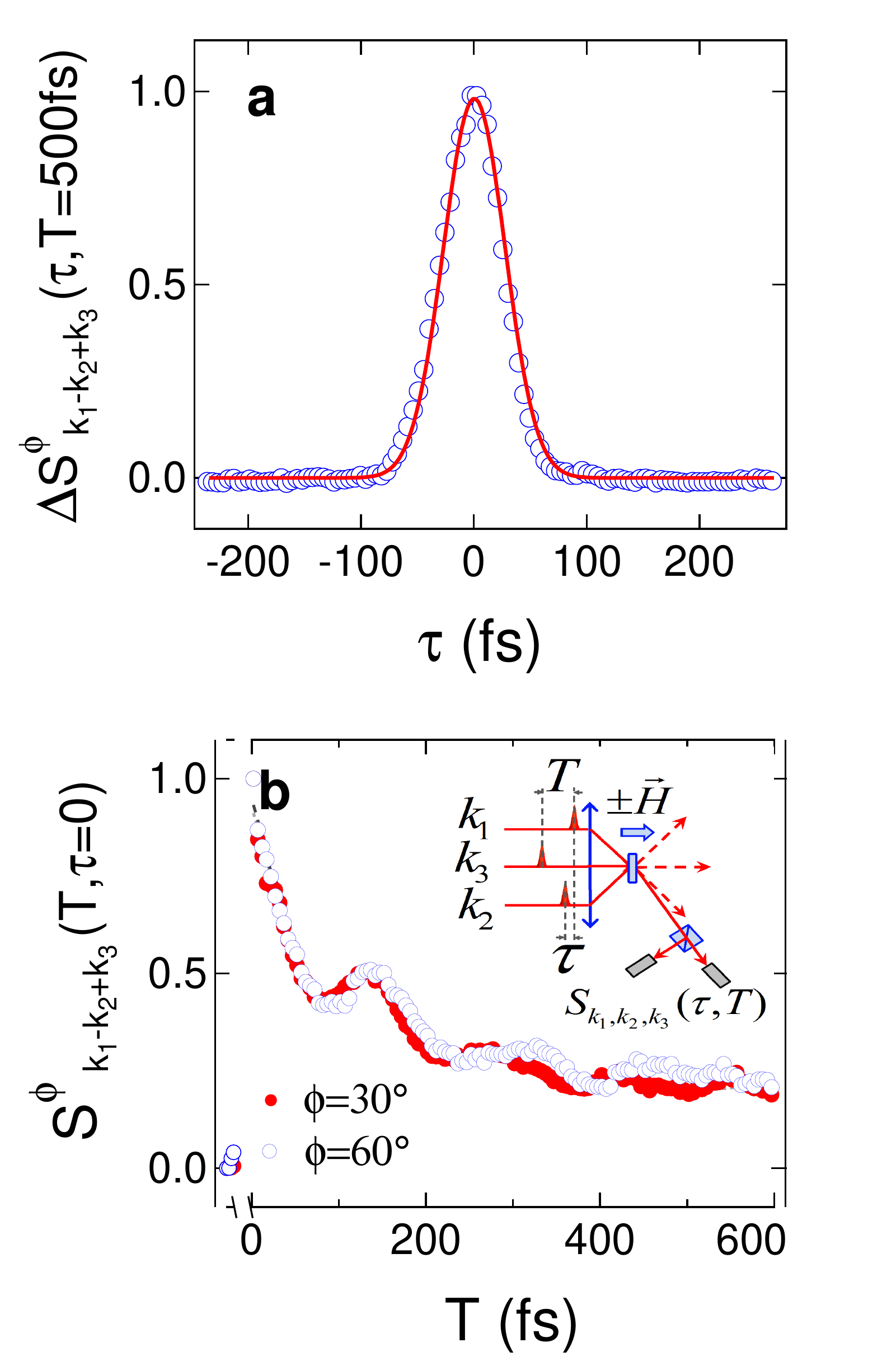}
\caption{\label{fig3}MO-FWM signal for a three-pulse configuration. a) Coherent response as function of $\tau$ for a delay $T = 500 fs$. b) Spin population dynamics as a function of T for a delay $\tau = 0$, showing the precession of the magnetization. Angle of magnetic field $\mathbf{H}$ : $\phi = 60^\circ$ (opened circles) and $\phi = 30^\circ$ (closed circles) and periods of precession: $T_p=\text{162 ps}$ and $T_p=\text{146 ps}$. Inset : three beams MO-FWM configuration.}
\end{figure}

The coherent spin-photon interaction is further confirmed experimentally in the three pulses MO-FWM configuration. As seen in figure 3, in that case, we separate the dephasing (Fig. 3a) and the population dynamics (Fig. 3b) of the spins states. Clearly the coherent response $\Delta S_{k_1-k_2+k_3}^\phi(\tau,T = \text{500 fs})$ is still dominated by the pulse duration (50 fs in that set of measurements). Conversely, the population dynamics $S_{k_1-k_2+k_3}^\phi(\tau = 0,T)$ contains the full spin population dynamics. It includes: 1) the thermalization of the spins, 2) the electron/spin-lattice relaxation $T_{1,spin-lat} \approx \text{1.35 ps}$ (not shown on this time scale), 3) the precession of the magnetization which we measured for two angles of the external magnetic field and 4) the heat diffusion with a time constant $T_{1,diff} > \text{1 ns}$. In Fig. 3b) the display focuses on the precession dynamics. The oscillations correspond to the projection of the motion of precession onto the magneto-optical polar direction, perpendicular to the sample plane. The precession period $T_p$ decreases with the field angle $\phi$ ($T_p(\phi = 60^\circ) = \text{162 ps}$;  $T_p(\phi = 30^\circ) = \text{146 ps}$). This is due to a change in the effective field which here is opposite to Cobalt thin films \cite{JYB2005} because this Bi-doped Garnet has a perpendicular anisotropy. The damping of the precession is fast $T_{1,prec} \approx \text{150 ps}$ and dominated by spin-phonon scattering and spatial inhomogeneities. Note that such precession dynamics is consistent with preceding works reported on the magneto-optical pump-probe dynamics, in the case of  transition metals \cite{Ju1999,VanKampen2002,Vomir2005} or ferrimagnetic Garnet \cite{Hansteen2005}. In that latter case it was also shown that one can manipulate the precession with temporal sequences of circularly polarized pump and probe pulses. The overall precession can be described by a three-temperature model for the electrons, the spins and the lattice as well as a modified Landau-Lifschitz-Gilbert equation that takes into account the non-conservation of the modulus as performed for cobalt thin films \cite{JYB2005}. The observed dynamics is well explained with our model of coherent coupling between photons and spins, including the spin-orbit interaction. Figure 4 shows the dephasing and population dynamics of the three pulses MO-FWM, using the hydrogen-like system \cite{Vonesch2012}. The parameters are: $T_1 = 1.35 ps$, $T_2 = 10 fs$, pulse duration: $50 fs$. For simplification, a single population life time is considered.

\begin{figure}
\includegraphics[scale=0.3]{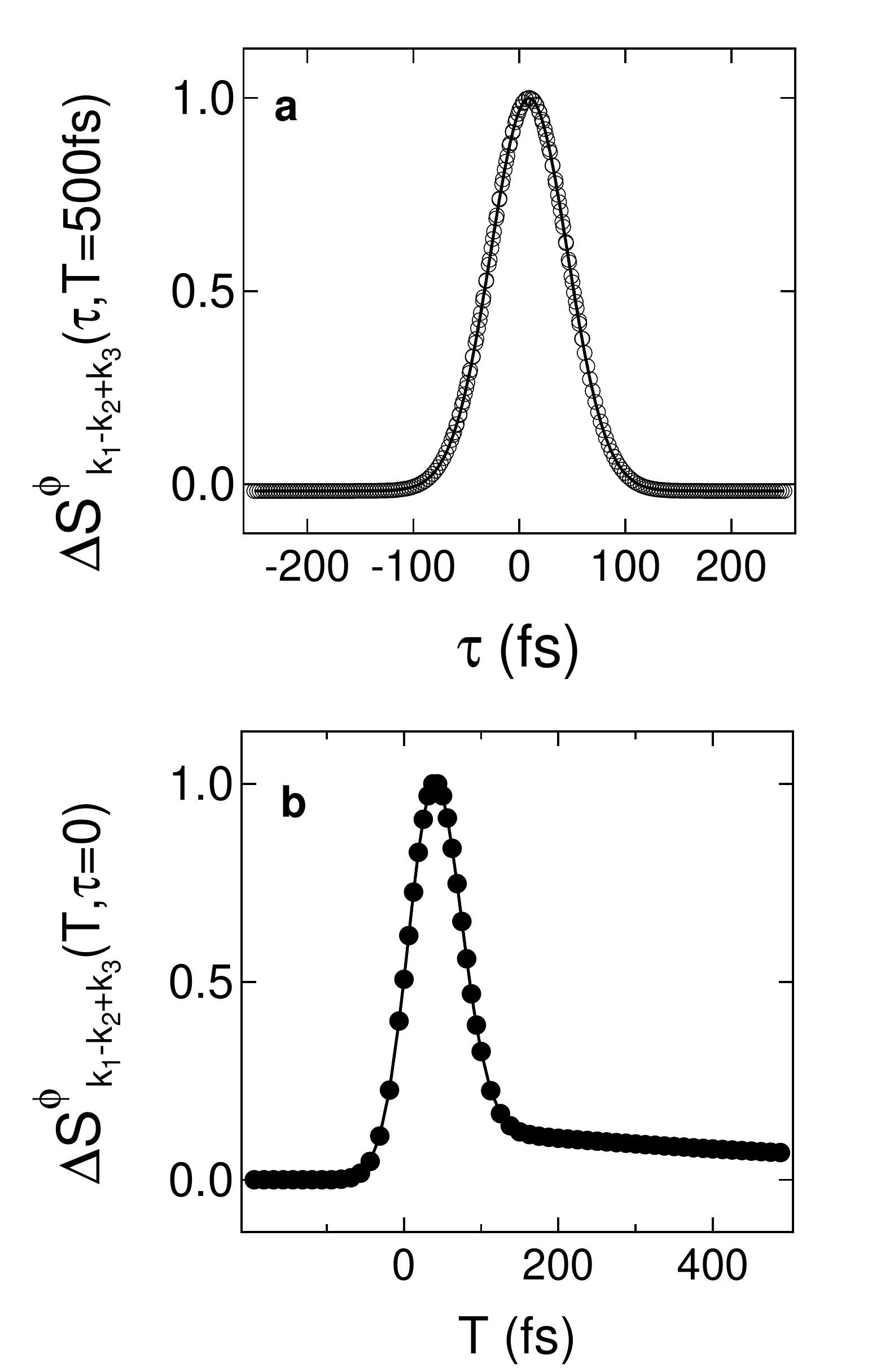}
\caption{\label{fig4}Model of MO-FWM for a three-pulse configuration. a) Coherent response as function of $\tau$ for a fixed delay $T = 500 fs$. b) Spin population dynamics as a function of T for a fixed delay $\tau = 0$.}
\end{figure}

In conclusion, the measurements of the magneto-optical four-wave mixing signals in a Bi doped Garnet film clearly reveal the spin-photon coherent coupling. The dephasing of the spins states occurs in a time scale much shorter than the populations relaxation which accounts for the magnetization dynamics usually observed in femtomagnetism. Our observations are well described by a model including the spin-orbit Hamiltonian applied to a simple 8-levels hydrogen-like system, with phenomenological dephasing time $T_2$ and lifetime $T_1$ of the spins states. Apart from understanding the initial photon/spin interaction in ultrafast magnetism, the experiments show the interesting potential of coherent magneto-optics for making femtosecond diffractive optical devices controlled by magnetic fields.

This work has been funded by the European Research Council under the project ERC-2009-AdG-20090325 $\sharp$247452. It has been partially supported by the Agence Nationale de la Recherche in France Equipex $\text{UNION ANR-10-EQPX-52-01}$.

\bibliography{Ref}


\end{document}